\documentclass[aps,prl,twocolumn,unsortedaddress,superscriptaddress]{revtex4-1}
\usepackage{graphicx,amssymb,amsmath,latexsym,epsfig}
\usepackage{color}

\newcommand{\cop}[2]{\hat{#1}^\dagger_{#2}}
\newcommand{\aop}[2]{\hat{#1}_{#2}}

\newcommand{\p}{\mathbf{p}}
\newcommand{\x}{\mathbf{x}}

\newcommand{\rr}{\mathbf{r}}

\begin{document}
\title{Kaluza-Klein tower of masses in compactified Bose-Einstein condensates
}

\author{Shay Leizerovitch}
\affiliation{School of Physics and Astronomy, Raymond and Beverly Sackler Faculty of Exact Sciences, Tel Aviv University, Tel-Aviv 69978, Israel.}
\author{Benni Reznik}
\affiliation{School of Physics and Astronomy, Raymond and Beverly Sackler Faculty of Exact Sciences, Tel Aviv University, Tel-Aviv 69978, Israel.}

\vskip.3in

\begin{abstract}
We propose a method for using ultracold atomic Bose-Einstein condensates, to form an analog model of a scalar field in five dimensional space-time, where one of the spatial dimensions is compact. In the analog system the extra dimension is discrete, and realized using the internal degrees of freedom of the system. In the low energy regime, the Kaluza-Klein tower of masses is recovered, and the effective radius of the compact dimension is identified.
\end{abstract}
\maketitle


Kaluza and Klein \cite{kaluza1921unitatsproblem,Klein1926}  (KK) proposed a simple ingenious method for incorporating electromagnetism into General Relativity  by adding an extra compact dimension to our ``ordinary" four dimensional space-time. In such a theory, electromagnetism arises naturally from the five dimensional Einstein-Hilbert action of General Relativity, and the electromagnetic fields become essentially geometrical objects.  The electron charge is derivable from first principles, and  expressed in terms of the five dimensional Newton's constant and the radius of the compact dimension. Since the momentum of matter fields that travel in a KK geometry is quantized along the compact dimension, the fields' masses become quantized as well- a phenomena known as the KK tower of masses 
\cite{bailin1987kk,overduin1997}. 

The KK theory was later-on extended to non abelian Yang-Mills gauge fields, and widely used in various proposals in attempt to go beyond the Standard Model of high energy physics (HEP).  In particular, KK compactification has been used in theoretical attempts to resolve the Hierarchy problem \cite{hamed1998,randall1999}, the cosmological constant problem \cite{rubakov1983kk-cos,mashhoon1994kk-cos}, the missing dark matter \cite{KK-DM}, and in  cosmological models of the early universe \cite{freund1982kk-cos,mashhoon1994kk-cos,dienes2017kk-eu}.

In the simplest case, one considers a five dimensional space-time with a ${\rm M}^4\times S^1$ topology, coordinates $x_{\alpha}=(x_\mu,y)$ where $\mu=0,1,2,3$, and a metric
$g^{(5)}_{\alpha\beta}$ defined by the line element
\begin{eqnarray}
(ds^{(5)})^2&=&\eta_{\alpha\beta}^{(5)}d x^\alpha d x^\beta\\
&=&\eta_{\mu\nu}d x^\mu d x^\nu - (dy+ q A_\mu(x) dx^\mu)^2~\nonumber 
\end{eqnarray}
with $\eta_{\mu\nu}=\text{diag}\lbrace 1,-1,-1,-1\rbrace$ ($c\equiv 1$).
Then, for a free (vacuum) five dimensional theory  defined by the Einstein-Hilbert  action, $S^{(5)}\propto \int  \sqrt{-g^{(5)}} R(g^{(5)}_{\alpha\beta}) d^5 x$. Integration over $y$ reduces $ S^{(5)}\to S^{(4)}=S_{GR}^{(4)}+S_{EM}^{(4)}$, where the functions $A_\mu$ in $g^{(5)}_{\alpha\beta}$ play the role of an electromagnetic vector potential.

To the pure gravitation action, one can add matter fields. A complex massless scalar field, $\Phi(x,y)$, that lives in five dimensions is described by the action $S_m^{(5)} = \int \sqrt{-g^{(5)}} \partial^\alpha\Phi \partial_\alpha\Phi^\dagger d^5x $. Since the matter field must also be periodic in the compact direction, $\Phi(x,y+2\pi r)=\Phi(x,y)$, (with $r$ being the radius of $S^{(1)}$) it has a Fourier decomposition
\begin{equation}\label{5d_field}
\Phi(x,y)=\sum_{n=-\infty}^{\infty} \phi_n(x)e^{i  {n y/r}}~.
\end{equation} 
Substituting the latter in $S_m^{(5)}$ and integrating over $y$, leads to the action $S_m^{(4)}=\sum_n\int d^{4}x [D^\mu\phi_n]^{\dagger}[D_\mu\phi_n]-n/r|\phi_n|^2$, with $D_\mu=\partial_\alpha + inq/r A_\mu$.
A low energy observer, who is insensitive to the extra compact structure of space-time, will therefore experience a set of charged fields coupled to a gauge field $A_\mu$, with charges $e_n=\pm nq/r$, and doubly degenerate masses $m_n^2=n^2/r^2$, along with a single free massless field for $n=0$. The latter eigenmass structure is known as the KK tower of masses.

Motivated by the success of atomic physics analog models in providing new insights into various quantum effects 
 in HEP and General Relativity  \cite{Barcelo2,Visser,Unruh,GarayPRA,Fedichev,Fischer,Barcelo2,BenniH,Reznik2000,Horstmann2010,Horstmann2011,Szpak}, by recent advances towards quantum HEP simulators of lattice gauge theory \cite{Cirac2010,ErezPrl,mazza2012optical,Zoller2012,ErezRevA,Erez2015,Wiese}, and in particular by recent proposals \cite{lewenstein2012,celi2015,carusottoQH4D,goldman2017} and experimental implementations \cite{celiScience2015,lensSience2015,spilmanSienec2015,lens2016,gadwaySience2017} for synthetic extra dimensions with atomic and photonic \cite{carusotto2016,carusotto2017} systems, we suggest and study a simple method that realizes a KK compactification, and manifests a tower mass spectrum  for (analog) relativistic massive fields, using atomic condensates.

It has been shown \cite{Visser}, that collective phonon excitations can behave, under certain conditions, as {\em massive} 
relativistic scalar fields, and more recently  \cite{leizerovitch2016},  that a scalar QED-like model - of {\em charged}  fields coupled to external electromagnetic fields -  has been obtained from doubly degenerate massive fields.

 Building upon these ideas, we shall here use a system with  $N$ trapped condensates, to construct a model that 
describes a massless field propagating on a manifold with a $M^4\times Z_N $ topology. ``Real" space-time dimensions are continuous, while the  extra compact dimension, built from the internal condensate states, is discrete.   
The system displays a mass spectrum that coincides, in the large $N$ limit, with that of a KK tower of massive relativistic fields in a compact $M^4\times S^1$ topology.

{\em Compactified BECs}. Let us consider a system with dilute ultracold alkali atoms with an odd number, $N$, of internal electronic levels that condensate and form an N-component BEC.   The interactions between the condensates' atoms, can be described as due to
s-wave scattering processes with corresponding magnitudes
 $U_{ij}$, $i,j = 1,..,N$.  We shall assume that self interactions, with $i=j$, are all equal: $U_{jj}=U=4\pi a/m$, where $a$ is the scattering length ($\hbar= 1$). 
The interactions between different atomic species arises from scatterings processes $U_{ij}$, $i\neq j$, as well as from external light fields induced couplings with homogeneous Rabi frequency $\Omega$. We shall consider a particular configuration wherein inter species interactions are  'compactified' along a periodic, chain-like structure, that has a rotational $Z_N$ symmetry.
Namely, a given specie $j$,  couples
only with 'nearest neighbour' condensates, $j-1$ and $j+1$, and  a periodic boundary conditions is imposed by identifying $j=N+1$ with $j=1$.
We further assume that $U_{ij}\equiv U^{\prime}$.

A Hamiltonian (density) that describes our system is given by
\begin{eqnarray}\label{Hamiltonian}
\mathcal{H}&=&\sum_{i,j=1}^{N}\delta_{ij}\left(\frac{1}{2m}\left|\nabla\psi_{i}(\x)\right|^{2}+\frac{U}{2}|\psi^{\dagger}_{i}(\x)\psi_{j}(\x)|^{2}\right)\\
&&+C_{ij}\left[\frac{U^{\prime}}{2}|\psi^{\dagger}_{i}(\x)\psi_{j}(\x)|^{2}+\Omega\left(\psi_{i}^{\dagger}(\x)\psi_{j}(\x)+\text{h.c}\right)\right],\nonumber
\end{eqnarray}
where $C_{ij}\equiv \delta_{i,j+1}+\delta_{i+1,j}$.

To study the effect of compactification on quasi-particle excitations, we use the 
mean-field approximation. We expand the condensates around their mean value, $\psi_i(\x)=\bar{\psi}_i+\delta\psi_i(\x)$, and assume equal uniform distributions, $\bar{\psi}_i=\sqrt{n}$.
In terms of the Fourier modes of the excitations $\delta\psi_i(\x)$,  
the Hamiltonian $\int \mathcal{H} d^3x $, reduces to a quadratic form.
To impose atom number conservation, we diagonalized the Gibbs energy functional using a Bogoliubov transformation \cite{Pethick2002,diagonalization1}.
This results with eigenmodes of the form  
\begin{equation}\label{excitation}
\varphi_j(\x)=\int \frac{d\p}{(2\pi)^{3}} e^{i\p \x}[u_{j}(\p)\aop{b}{j,\p}+v_{j}(-\p)\cop{b}{j,-\p}]
\end{equation}
with Bogoliubov amplitudes given by 
\begin{equation}
u_j(\p),v_j(-\p)=\pm\sqrt{\frac{1}{2N}\left(\frac{mc_s^{2}+p^{2}/2m}{E_j}\pm1\right)}~.
\end{equation}
The associated  energy spectrum is given by 
\begin{equation}\label{spectrum}
E_j=\sqrt{E_{rj}^{2}+c_{sj}^{2}p^{2}+(p^{2}/2m)^{2}}~,
\end{equation}
where
\begin{eqnarray}
E_{rj}^{2}&=&4\left[\Omega^2-nU\Omega+\left(nU\Omega-2nU^{\prime}\Omega-2\Omega^2\right)\cos \alpha_j\right.\nonumber\\
&&\left.+\left(2nU^{\prime}\Omega+\Omega^2\right)\cos ^2\alpha_j\right]~,\label{mass}\\
mc^{2}_{sj}&=&nU-2\Omega+2\left(nU^{\prime}+\Omega \right) \cos \alpha_j~.\label{speed}
\end{eqnarray}
Here $\alpha_j\equiv 2\pi j/N$, and $c_{sj}$ is the speed of sound of the jth mode.
As one can see, the spectrum \eqref{spectrum} differs from the case of a one component BEC. 

We observe that the interaction between the species give rise to an energy gap.
The lowest mode, $j=0$, behaves as a massless phonon, but the remaining N-1 modes are all gapped. We further notice that for $j\neq0$,  the levels $j$ and $N-j$ are degenerate. Hence the spectrum can be organized to $(N-1)/2$ doubly degenerate branches.
As we shall discuss in the following, each branch can be seen as describing a pair of massive relativistic scalar fields, or, equivalently, a single complex (``charged'') massive field. Moreover, fields of different branches have different sound velocities $c_{sj}$.
Rather than having a single common metric, each complex field experiences a different (effective) causal structure - or as referred to in  \cite{Visser} we have \emph{multi metric} structure.

\vspace{.1cm}
{\em  Tower of masses}.
 The speed of sound introduces a characteristic energy (and length) scale, that divides the spectrum into two regimes;
In the high momentum regime, $p\gg \sqrt{2}mc_s$, free excitations possess a non-relativistic, quadratic dispersion relation, and since we wish to simulate a relativistic field theory, we have less interest in it.
Nevertheless we will present the resulting spectrum for this limit and briefly discuss it.
In the low momentum regime, $p\ll \sqrt{2}mc_s$, the spectrum of a gapless mode is approximately linear and the modes have a ``collective" nature \cite{pitaevskii2003}.
In this regime BEC excitations are analogous to a relativistic massless scalar field, exhibiting a ``causal" sound cone determined by the speed of sound $c_s$.
A  massive scalar field analog can also be realized \cite{Visser}, if the energy gap is properly bounded with respect to the characteristic energy scale $mc_s^{2}$, by tuning the coupling laser Rabi frequency \cite{leizerovitch2016}.
This effective low energy behavior is a direct result of the non linear interaction term which appears in the GP Hamiltonian \eqref{Hamiltonian}, while discarding it leads to a trivial non-relativistic dispersion. This is also true for the Bose-Hubbard model \cite{lewenstein2012book}.

\vskip .3cm

Observing Eq. \eqref{speed}, we conclude that there are $(N+1)/2$ distinct sound cones, where each pair of eigenmodes, $j$ and $N-j$ (for $j>0$), acquires different propagation speed.
Given that, the simulating system's eigenmodes cannot be identified with the mass modes of the KK theory, since each eigenmode exhibits a different causal structure. Hence we cannot reassemble the five dimensional scalar field \eqref{5d_field}, and the analogy between the two theories becomes meaningless.
Nevertheless, we can compare the different sound propagation velocities $c_{sj}$ by introducing the \emph{mono-metricity} condition: $nU^{\prime}=-\Omega$.
The resulting speed of sound is
\begin{equation}
c_s=\sqrt{\frac{nU-2\Omega}{m}}=\sqrt{\frac{nU+2nU^{\prime}}{m}}~.
\end{equation}
This emphasize the significant role of the inter-species atomic interactions in simulating extra dimensional relativistic QFT, as it ``gauges" away the discrepancy in the different eigenmodes' causal structure.
\begin{figure}[t]
                \includegraphics[width=0.44\textwidth]{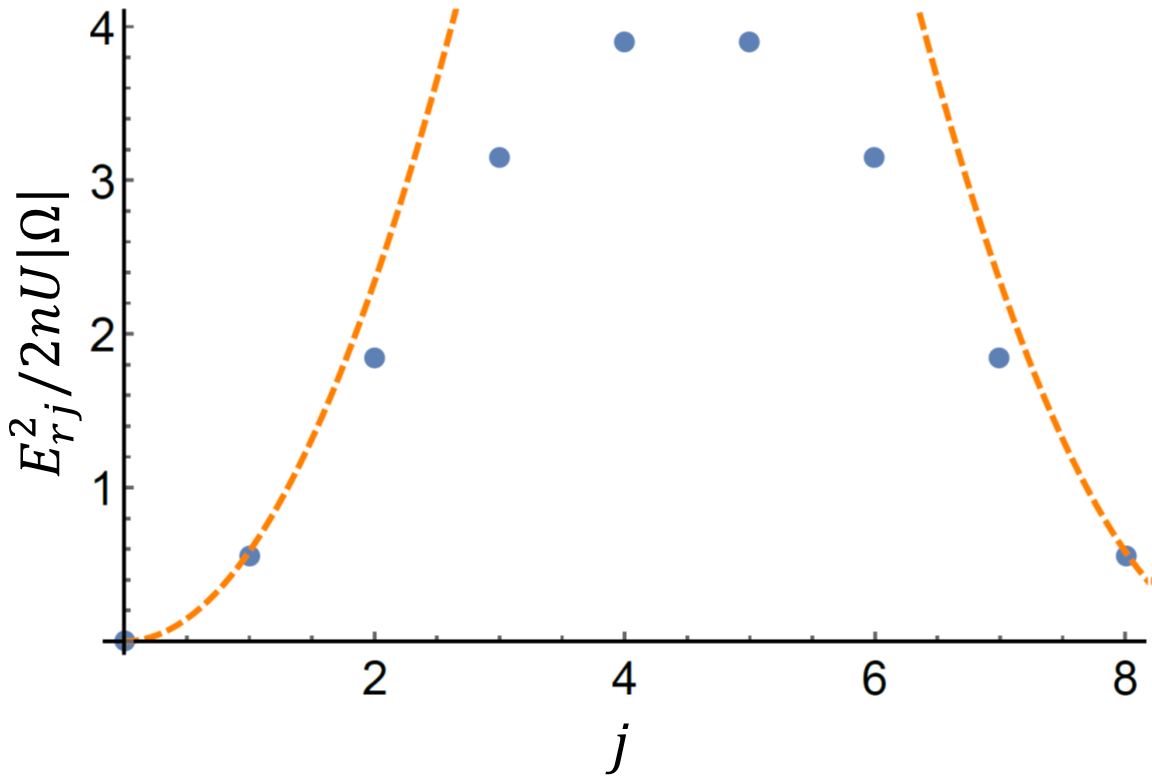}
                \caption{The squared rest mass for $N=9$ is given in natural units as a function of $j$, with $|\Omega|/nU=0.1$. The blue dots denote the mass eigenmodes \eqref{rest_mass}. The orange (dashed) curves correspond to the approximate rest mass \eqref{approx_mass}. There are five modes that coincide with the continuum limit. In terms of the simulated theory, the right branch modes correspond to $\phi_{-n}$.}
                \label{fig:rest_mass}
\end{figure}
\vskip .3cm

Substitution of the mono-metricity condition in the eigenmodes' rest mass \eqref{mass}, leads to the following expression:
\begin{equation}\label{rest_mass}
E_{rj}^{2}=-4\Omega\left[nU\left(1-\cos\alpha_j\right)-\Omega\sin^2\alpha_j\right]~.
\end{equation}
We observe that for $\Omega>0$ the effective mass is imaginary, and may lead to dynamical instability \cite{pitaevskii2003}. In the simulated theory perspective, the negative squared mass eigenmodes are tachyons, and the extra dimension has a timelike signature \cite{bailin1987kk}. In order to avoid it, we set $\Omega$ to be negative.

We can already identify a ``tower of masses", consisting of one massless mode ($\varphi_0$), and degenerate pairs ($\varphi_j$, $\varphi_{N-j}$).
But in order to reveal the full analogy between the two theories, we first need to apply the continuum limit $j\ll N$, resulting with
\begin{equation}\label{approx_mass}
E_{rj}^{2}\approx 2\left|\Omega\right|\left(nU+2\left|\Omega\right|\right)\left(\frac{2\pi  j}{N}\right)^{2}= c_s^{2}p_5^{2}~.
\end{equation}
Here $p_5\equiv 2\pi j/Na$ is the discrete fifth dimensional momentum, where we have identified $a\equiv (2m\Omega)^{-1/2}$ as the effective ``lattice spacing".
Given that, we readily obtain the radius of the synthetic dimension, $r=N/2\pi\sqrt{2m\Omega}$.
Since (for a gapless mode) a linear dispersion relation is obtained far below the cutoff energy scale $mc_s^{2}$, it follows that $E_{rj}$ should be bounded by the same energy scale. Observing Eq.\eqref{approx_mass}, it also follows that $p_5\sim p\ll \sqrt{2}mc_s$ and
\begin{equation}\label{momentum_constraint}
\frac{p_5}{\sqrt{2}mc_s}=\frac{2\pi j}{N}\sqrt{\frac{|\Omega|}{nU+2|\Omega|}}=2\pi j\frac{\xi}{Na}\ll 1~.
\end{equation}
Indeed, in the limit $N\rightarrow\infty$, the latter is satisfied for every finite $j$.
However, in realistic experimental systems, the number of condensed species is limited and may reach up to $N=10$. Hence the simulating system includes only several low energy modes, as depicted in Fig.\ref{fig:rest_mass}.
Demanding that \eqref{momentum_constraint} still holds, we must further require that $(2mL^{2})^{-1}<\Omega\ll nU$ where $L$ is the typical length of the system. Using the definition of the effective length $a$, we conclude that the analogy is valid if the length of the synthetic dimension is much larger than the healing length, $\xi\equiv (\sqrt{2}mc_s)^{-1}$.

Now that we have shown that the excitations' spectrum of a multi-component BEC coincides with that of the simulated theory in the long wavelength regime, we shall continue with the construction of the simulating scalar field $\tilde{\Phi}(x,\theta_{k})$.
We have shown in a previous paper \cite{leizerovitch2016} that a scalar field can be built from the imaginary part of a BEC excitation \eqref{excitation}, $\tilde{\phi}(\x)=\text{Im}\left\lbrace\varphi(\x)\right\rbrace/\sqrt{2}$. Using the latter relation, and the ``group velocity" in the compact dimension, we can identify the analog fields $\tilde{\phi}_j$ with the simulated field's mass modes $\phi_n$. For $1\leq j\leq (N-1)/2$, the fields $\tilde{\phi}_j$ are associated with mass modes with $n>0$, and the fields $\tilde{\phi}_{N-j}$, with mass modes with $n<0$. The two massless modes $\tilde{\phi}_0$ and $\phi_0$ are identified as well.
Finally, we can define the analog field
\begin{equation}\label{5d_simulating_field}
\tilde{\Phi}(x,y_{k})=\frac{1}{\sqrt{N}}\sum_{j=0}^{N-1}\tilde{\phi}_j(x)e^{-i p_5 y_k }~,
\end{equation}
where $y_k=ak$ is the position in the discrete synthetic dimension with $k=0,...,N-1$.

\begin{figure}[t]
                \includegraphics[width=0.45\textwidth]{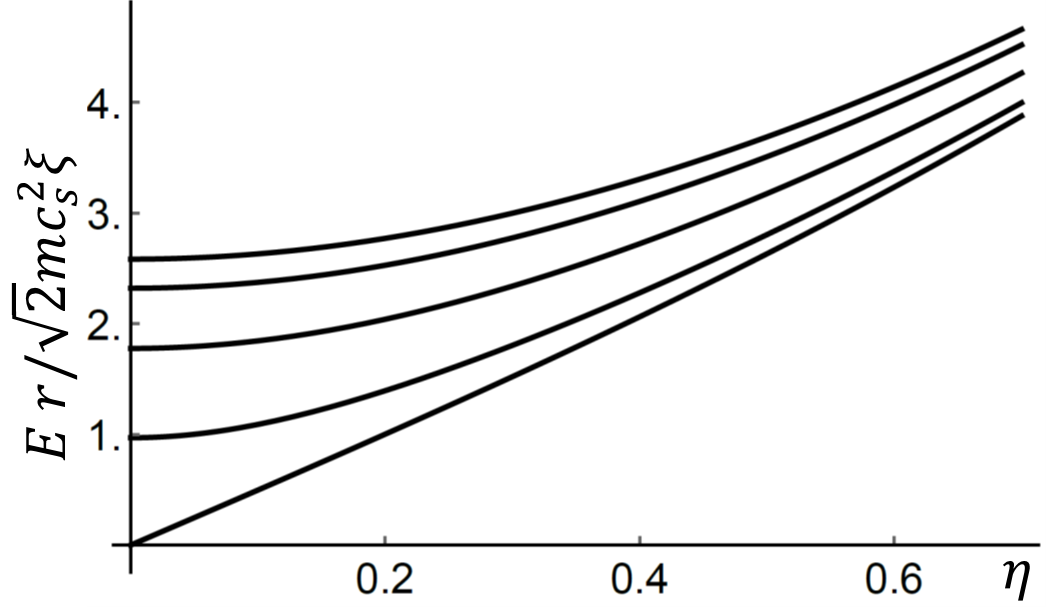}
                \caption{The dispersion relation \eqref{spectrum} plotted in natural units as a function of $\eta\equiv p\cdot\xi$ for $N=9$. The different curves represent different values of $j$. For $\eta\ll 1$, the spectrum coincide with the KG dispersion relation.}
                \label{fig:rel_dispersion}
\end{figure} 

\vspace{.3cm}
{\em Correlation functions}.
Next we discuss how the (effective) dimensionality can be revealed through correlation functions, which are experimentally measurable.
To this end, we shall employ the relations we made between the fields $\tilde{\phi}(x,y_k)$ and $\phi(x,y)$, and construct from them two point correlation functions.

The dimensionality of a system is one of the properties that define the scaling of a two-point correlation function. 
In the theory we wish to simulate, the correlation function between the two points $\rr_1=(\x_1,y_1)$ and $\rr_2=(\x_2,y_2)$, is
\begin{equation}\label{simulated_corr}
D(s,\Delta)= \frac{1}{2\sqrt{2}\pi^{2}\xi^{3}}\frac{R_l}{(s^{2}+(R_l \Delta)^{2})^{3/2}}
\end{equation}
where $s\equiv|\x_2-\x_1|/\xi$, $\Delta\equiv|k_2-k_1|$ and $R_l=a/\xi$.

\begin{figure}[t]
                \includegraphics[width=0.45\textwidth]{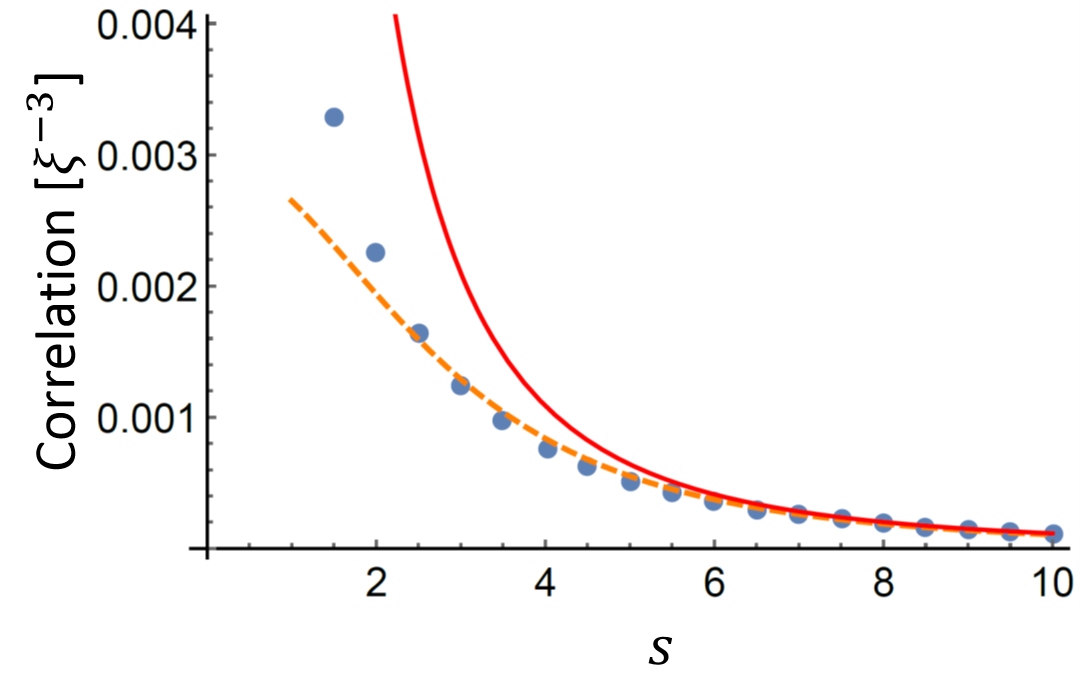}
                \caption{The correlation functions, given in units of $\xi^{-3}$ as a function of $s$ for $\Delta=1$, and $a/\xi\gg 1$. The analog field's correlator $\tilde{D}(s,1)$ \eqref{simulated_corr} is denoted by blue dots, $D(s,1)$ \eqref{simulating_corr} denoted by a dashed orange line, and $\tilde{D}_{tr}(s,1)$ \eqref{trun_corr} denoted by a red solid line. For $s>10$, the three coincide.}
                \label{fig:correlation}
\end{figure}

\noindent We therefore wish to examine if, and under which conditions, the two point correlation function of the analog field coincides with this result.
We have calculated numerically the correlator of the analog field \eqref{5d_simulating_field} for $N=9$
\begin{equation}\label{simulating_corr}
\tilde{D}(s,\Delta)=\sum_{j=1}^{9}e^{-i\frac{2\pi}{N}j\Delta}\int\frac{d^{3}\mathbf{\eta} e^{-i\mathbf{\eta}\,s}}{(2\pi\xi)^{3}}(u_j(\mathbf{\eta})-v_j(\mathbf{\eta}))^{2},
\end{equation}
in the limit $Na\gg \xi$, where $\eta=\xi p$.
The result is plotted in Fig.\ref{fig:correlation} for $\Delta=1$ (neighbouring sites) as a function of $s\equiv \Delta x/\xi$, and denoted by blue dots.
Comparing it to $D(s,\Delta)$ (denoted by an orange dashed line), we conclude that the two coincide for a distance much larger than the healing length ($s>10$).

It is also interesting to examine whether in the relativistic limit, $\tilde{D}(|r_{12}|)$ also agrees with $D(|r_{12}|)$ , and if it is a solid approximation.
Namely, we consider a truncated correlator that consists from the low energy modes for which the continuum limit applies, $|j|\leq j_{tr}=2$ for $N=9$ (see Fig.\ref{fig:rest_mass}), and apply the relativistic limit $(u_j(\p)-v_j(\p))^{2}\approx 2mc_s^{2}/E$:
\begin{equation}\label{trun_corr}
\tilde{D}_{tr}(s,\Delta)=\frac{1}{9}\sum_{j=-j_{tr}}^{j=j_{tr}}\frac{R_{m}(j) K_1[R_{m}(j)s]}{\sqrt{2} \pi ^2 s}
\end{equation}
where $R_{m}(j)\equiv R_l^{-1}\alpha_j$, and $K$ is the modified Bessel function of the second kind.
The latter is also plotted in Fig.\ref{fig:correlation}, denoted by a solid red line.
As one can see, for $s>10$ the three correlation functions merge.

\vspace{.3cm}

{\em Non-relativistic limit}.
When the length of the synthetic dimension is in the order of the healing length $Na\sim \xi$,  or smaller, the model has a non-relativistic spectrum. 
In this case we do not impose the mono-metricity condition, but demanded that $\Omega\gg nU,nU^{\prime}$.
We then obtain the following non-relativistic spectrum
\begin{equation}\label{non_rel_dispersion}
E\approx p^{2}/2m+2\Omega(1-\cos\alpha_j)+n(U+U^{\prime}\cos\alpha_j)~.
\end{equation}
As in the previous case, in the limit $j\ll N$, we obtain also here a tower of masses \cite{lewenstein2012}.

\vskip .3cm

{\em Discussion}. In this letter, we suggested and studied a method for obtaining an effective extra dimension using Bose Einstein condensates, that gives rise to the KK structure of  a mass tower spectrum for a relativistic scalar field. As the analog field is composed, the dimensionality of the system can be tested experimentally by measuring the correlation functions.
The method does not require single addressing of atoms; a compact structure of the extra dimension is obtained by tailoring the interactions between the condensates' components. The KK tower of masses is then obtained by demanding a mono-metric structure, and applying the phononic limit for $p_5$. Both fulfilled by controlling a single parameter- the Rabi frequency of external laser.
The last requirement of a mono-metric causal structure is not needed in the non-relativistic limit of the model.

\bibliography{refs}
\end{document}